\documentclass[twocolumn]{aastex62}

\usepackage{xspace}
\usepackage{booktabs}
\usepackage{multirow}

\newcommand{\V}[1]{\mathbf{#1}}
\newcommand{\bra}[1]{\left ( #1 \right )}

\newcommand{\mean}[1]{\left < #1 \right >}
\newcommand{\Ms}{\mathrm{M_s}}
\newcommand{\Ma}{\mathrm{M_a}}
\newcommand{\pd}{\partial}
\newcommand{\pb}{\beta_{\mathrm{p}}}
\newcommand{\cs}{c_{\mathrm{s}}}
\newcommand{\vA}{v_{\mathrm{A}}}

\newcommand{\pth}{p_\mathrm{th}}
\newcommand{\pB}{p_\mathrm{B}}
\newcommand{\ptot}{p_\mathrm{tot}}
\newcommand{\PTI}{$\mathtt{M0.70P0.37}_\mathrm{Cool: ff}^{\gamma=5/3}\mathtt{A1.00H}$\xspace}
\newcommand{\PMS}{$\mathtt{M0.54P0.58}_\mathrm{Cool: ff}^{\gamma=5/3}\mathtt{A1.00H}$\xspace}

\newcommand{\Skew}{\mathrm{skew}}
\newcommand{\Kurt}{\mathrm{kurt}}

\newcommand{\Ccool}{C_\mathrm{cool}}

\newcommand{\eqref}[1]{(\ref{#1})}

\begin{document}
\title{As A Matter of State: The role of thermodynamics in magnetohydrodynamic turbulence}
\shorttitle{Thermodynamics in MHD turbulence}
\shortauthors{Grete, O'Shea \& Beckwith}
\correspondingauthor{Philipp Grete}
\email{grete@pa.msu.edu.}

\author[0000-0003-3555-9886]{Philipp Grete}
\affiliation{
Department of Physics and Astronomy,
Michigan State University, East Lansing, MI 48824, USA}
\author{Brian W. O'Shea}%
\affiliation{
Department of Physics and Astronomy,
Michigan State University, East Lansing, MI 48824, USA}
\affiliation{
Department of Computational Mathematics, Science and Engineering,
Michigan State University, East Lansing, MI 48824, USA}
\affiliation{
National Superconducting Cyclotron Laboratory,
Michigan State University, East Lansing, MI 48824, USA}

\author{Kris Beckwith}
\affiliation{%
Sandia National Laboratories, Albuquerque, NM 87185-1189, USA
}

\keywords{MHD --- methods: numerical --- turbulence}

\begin{abstract}
Turbulence simulations play a key role in advancing the general
understanding of the physical properties turbulence and in interpreting astrophysical observations of turbulent plasmas.
For the sake of simplicity, however,  turbulence simulations are often conducted
in the isothermal limit.
Given that the majority of astrophysical systems are not governed by isothermal dynamics,
we aim to quantify the impact of thermodynamics on the physics of turbulence, through varying adiabatic index, $\gamma$, combined with a range of optically thin cooling functions.
In this paper, we present a suite of ideal magnetohydrodynamics simulations of thermally balanced stationary turbulence in the subsonic, super-Alfv\'enic, high $\pb$ 
(ratio of thermal to magnetic pressure) regime, where turbulent dissipation is balanced by two idealized cooling functions (approximating
linear cooling and free-free emission) and examine the impact of the equation of state by considering cases that correspond to isothermal, monatomic and diatomic gases.
We find a strong anticorrelation between thermal and magnetic pressure independent of
thermodynamics, whereas the strong anticorrelation between density and magnetic field 
found in the isothermal case weakens with increasing $\gamma$.
Similarly, with the linear relation between variations in density and thermal 
pressure with sonic Mach number becomes steeper with increasing $\gamma$.
This suggests that there exists a degeneracy in these relations with respect to 
thermodynamics and Mach number in this regime, which is
dominated by slow magnetosonic modes. These results have implications for attempts to infer (e.g.) Mach numbers from (e.g.) Faraday rotation measurements, without additional information regarding the thermodynamics of the plasma. However, our results suggest that this degeneracy can be broken by utilizing higher-order moments of observable distribution functions.

\end{abstract}

\section{Introduction}
Magnetic fields are ubiquitous in the Universe and have been observed
on all scales, from stellar and planetary systems to the intracluster medium.
Similarly, many astrophysical systems are expected to be governed by or
subject to turbulence simply by the large spatial scales involved \citep{Brandenburg2013}.
More generally, magnetized turbulence is thought to play a key role in many 
astrophysical systems and processes, e.g., magnetic field amplification via the turbulent 
dynamo \citep{Tobias2013,Federrath2016a}, particle acceleration in shock fronts resulting
in cosmic rays \citep{Brunetti2015}, or the formation of jets \citep{Beckwith2008} and in accretion disks \citep{Balbus1998}.

In the absence of detailed 3D spatio-temporal observations and/or experimental data,
numerical simulations are often used to support the interpretation of observations or, in 
the case of turbulence research, have become one of the major drivers
of scientific advances.
This pertains, for example, to studying energy dissipation 
and turbulent energy cascades, which serves to illuminate the physical mechanisms of energy redistribution and and the local nature of energy transfer within turbulence
\citep[e.g.,][]{Yang2016,Grete2017a,Andres2018} or to turbulence modeling
\citep{Clark1979,Germano1991,Chernyshov2012,Grete2016a}, which allows the
incorporation of small-scale turbulent effects and feedback in turbulence simulations that usually are
not able to capture the full dynamical range.

From an astrophysical point of view, a range of studies have analyzed turbulence dynamics and 
statistics in a variety of regimes, with a focus on quantities related to observations enabling
inference of statistical properties of the plasma below the observational resolution limit.
In particular, in the star formation community significant attention is paid to the
relation between density fluctuations and sonic Mach number ($\Ms$).
In the isothermal, supersonic case, the density distribution is well-described
by a lognormal distribution and the width of the distribution proportional to the
sonic Mach number \citep{Padoan1997,Passot1998}.
Moreover, for a given $\Ms$, the standard deviation of the density fluctuations 
contains information on the effective turbulent production mechanism, with respect to (for example)
the effects of different ratios of compressive to rotational modes in the 
forcing \citep{Federrath2008}.
Similarly, the departure from an isothermal equation of state (EOS) has been studied for 
hydrodynamic turbulence and for a polytropic EOS \citep{Federrath2015a} or
an adiabatic EOS \citep{Nolan2015,Mohapatra2019} indicating additional dependencies, on (for example) the adiabatic index $\gamma$, the relation between density 
fluctuations and $\Ms$.
In the magnetized case the majority of studies focus on the isothermal case,
finding an additional dependency on the ratio of thermal to magnetic pressure
$\pb$ \citep[e.g.,][]{Kowal2007,Padoan2011,Molina2012}.
Overall, a highly dynamic picture of turbulence has been found, challenging our ability to examine the general case.

In this paper, we present adiabatic magnetohydrodynamic simulations of stationary 
turbulence in the subsonic ($\Ms \approx 0.2$ to $0.6$),
super-Alfv\'enic ($\Ma \approx 1.8$), and high $\pb$ (ratio of thermal to magnetic 
pressure with $10 \lesssim \pb \lesssim 100$) regime.
This regime approximates the turbulent intracluster medium (ICM)
\citep{Brunetti2007,Bruggen2015} even though our MHD model neglects effects stemming
from low collisionality \citep{Schekochihin2006,Schekochihin2008}.
However, studies including effects from low collisionality, e.g., through the 
Chew–Goldberger–Low MHD model, found that they generally  introduce only small differences
compared to the MHD model, e.g., a small increase in density fluctuations
\citep{Kowal2011,Santos-Lima2017}, but leave imprints on Faraday rotation maps
\citep{Nakwacki2016}.
Here, we specifically focus on the effects of departure from an isothermal equation of state by
varying the adiabatic index $\gamma$ and the cooling function between simulations.

The rest of this paper is organized as follows.
In Sec.~\ref{sec:method}, we introduce the numerical setup and the simulations conducted.
In Sec.~\ref{sec:results}, we present results from analyzing the simulations starting
with a high-level overview of the energy spectra,
to correlations in the 
stationary regime, to statistics of distribution functions that can potentially 
be used to break degeneracies between the thermodynamics and sonic Mach number.
The results are discussed in Sec.~\ref{sec:disc}, and we conclude with a brief outlook in Sec.~\ref{sec:conclusions} 
on how our findings point the way to future measurements that can be used to better diagnose the properties of turbulence in astrophysical plasmas.

\section{Numerical Details}
\label{sec:method}
In this work we utilize the equations of compressible, ideal magnetohydrodynamics (MHD)
equations
\begin{eqnarray}
\label{eq:rho}
\pd_t \rho +  \nabla \cdot \bra{\rho \V{u}} = 0 \,, \\
\label{eq:rhoU}
\pd_t \rho \V{u} + \nabla \cdot \bra{ \rho \V{u} \otimes \V{u} -  \V{B} \otimes \V{B}
+ \mathcal{I} \ptot} = \rho \V{a} \,,  \\
\label{eq:B}
\pd_t \V{B} -  \nabla \times \bra{ \V{u} \times \V{B}}  = 0 \,,  \\
\pd_t E + \nabla \cdot \bra{\bra{E + \ptot}\V{u} - \V{B}\bra{\V{B}\cdot\V{u}}} = 
\rho\V{a}\cdot\V{u}- \mathcal{L} \,, 
\end{eqnarray}
that are closed by an ideal equation of state $\pth = \bra{\gamma-1}\rho e$,
with $\gamma$ as the ratio of specific heats.
The symbols have their usual meaning, i.e., density~$\rho$, velocity~$\V{u}$,
total pressure $\ptot = \pth + \pB$ consisting of thermal pressure~$\pth$
and magnetic pressure $\pB = 1/2 B^2$, 
and magnetic field~$\V{B}$, which includes a factor~$1/\sqrt{4\pi}$.
Cooling is included via $\mathcal{L}$ and 
$E = 1/2 \bra{\rho u^2 + B^2} + \rho e$ is the total energy density with 
specific internal energy $e$.
Vector quantities that are not in boldface refer to the $L^2$ norm of the vector and
$\otimes$ denotes the outer product.
The details of the acceleration field~$\V{a}$ that we use to mechanically drive our 
simulations are described in Section~\ref{sec:sim-overview}.

\subsection{Cooling functions}

The cooling curve for optically thin, astrophysically relevant plasmas
is not scale-free, and as such its use is undesirable if we wish to
achieve an understanding of non-isothermal turbulence in a broader
context.  To that end, we use two idealized cooling functions in this work:
linear cooling with
\begin{equation}
\mathcal{L} = \Ccool \rho e  \propto \rho e \,,
\end{equation}
and cooling that approximates free-free emission with
\begin{equation}
\mathcal{L} = \Ccool \rho^2 e^{1/2} \propto \rho^2 e^{1/2} \,.
\end{equation}
In this idealized setup appropriate units are absorbed in $\Ccool$.
In the majority of the simulations $\Ccool$ is chosen to approximately
balance turbulent dissipation (which, in turn, balances the energy injection 
from the forcing).
In the case of linear cooling
\begin{equation}
\mean{\rho} \mathrm{U}^3/\mathrm{L} 
\approx \mean{\mathcal{L}} \approx \Ccool \mean{\rho} \mean{e}
\approx \Ccool \mean{\pth}/\bra{\gamma-1}\,,
\end{equation}
where the means, $\mean{\cdot}$, refer to the spatial mean value in the
stationary regime, $U$ is the root mean square velocity in the
simulation's stationary regime, and $L$ is the characteristic
turbulence length scale.

\subsection{Implementation}
\label{sec:numerics}
All simulations were conducted with a modified version of 
the astrophysical MHD code \texttt{Athena 4.2}
\citep{Stone2008} 
using the same numerical scheme consisting of
second order reconstruction with slope-limiting 
in the primitive variables, an HLLD Riemann solver, 
constrained transport for the 
magnetic field, and a MUSCL-Hancock integrator \citep{Stone2009}.
Moreover, we used first order flux correction \citep{Lemaster2009,Beckwith2011}
in cells where the second order scheme described above results in a negative density 
or pressure -- the integration is repeated using first order reconstruction.
Explicit viscosity and resistivity are not included and thus
dissipation is of a numerical
nature given the shock capturing finite volume scheme, making the simulations
implicit large eddy simulations \citep{grinstein2007implicit}.
Strictly speaking the equations governing the simulations are not the ideal 
MHD equations but include implicit dissipative terms. The nature of implicit dissipation in the
\texttt{Athena} code was examined by \cite{Simon2009,Salvesen2014}  \cite[see also][]{Beckwith2019};
the work of these authors demonstrated the similarity of these terms to explicit viscosity and resistivity, similar to techniques adopted for inviscid hydrodynamics \citep{Sytine2000}.

In order to achieve a stationary regime with constant Mach number in a
driven, adiabatic simulation a mechanism to remove the dissipated
energy is required.
We implemented a flexible cooling mechanism
\footnote{
All modifications are available in our fork at 
\url{https://github.com/pgrete/Athena-Cversion}.
The simulations were run with changeset \texttt{3a7c300}.
}
approximating optically thin cooling.
In addition to the cooling function itself, we added several constraints
to the integration cycle.

First, the timestep is limited so that the internal energy is not changing by
more than 10\% per cycle.

Second, a cooling floor is employed in the form of a pressure floor.
For all simulations cooling is turned off in a given cell during a
given time step if the pressure drops to values of 
less than $10^{-4}$ in code units (which is $\simeq 10^{-4}$ of the
mean initial value in the calculations).

Third, we ported the ``entropy fix'' of \citet{Beckwith2011} for relativistic
MHD to the non-relativistic case.
The entropy fix introduces the entropy as a passive scalar to the set of equations
solved.
In case the first order flux correction fails, i.e., if density or thermal pressure are still 
negative in a cell after a first order update, the entropy is used to recover 
positive values.
However, this fix was only required in the two simulation that were thermally 
marginally stable (\PMS) and thermally unstable (\PTI),
see Sec.~\ref{sec:pdfs}, and even in those cases only 
tens out of $1024^3$ cells were affected per simulation for a very
small fraction the timesteps.

\subsection{Simulations}
\label{sec:sim-overview}
\begin{table*}
\begin{center}
\begin{tabular}{lrlllrrrrrr}
\toprule
& \multicolumn{7}{c}{Simulation/initial parameters} & \multicolumn{3}{c}{Stationary regime} 
\\
\cmidrule(lr){2-8} 
\cmidrule(lr){9-11} 
 ID                                                 &   Resolution & $\gamma$   & Cooling   & $\Ccool$   &   $a$ &   $p_\mathrm{th,init}$ &   $\beta_\mathrm{p,init}$ &    $\Ms$ &    $\Ma$ &   $\pb$ \\
\midrule
$\mathtt{M0.23P1.00}\mathtt{iso}\mathtt{A0.25}$                       &     $512^3$ & 1          & no        & --                  &  0.25 &                   1.00 &                     290 &    0.23(1) &    1.93(9) &    141(6) \\
 $\mathtt{M0.37P1.00}\mathtt{iso}\mathtt{A0.56}$                       &     $512^3$ & 1          & no        & --                  &  0.56 &                   1.00 &                      71 &    0.37(1) &    1.64(6) &     44.7(1.5) \\
 $\mathtt{M0.50P1.00}\mathtt{iso}\mathtt{A1.00H}$                       &    $1024^3$ & 1          & no        & --                  &  1.00 &                   1.00 &                      71 &    0.50(1) &    1.59(5) &     22.4(1.5) \\
 $\mathtt{M0.59P0.74}\mathtt{iso}\mathtt{A1.00}$                       &     $512^3$ & 1          & no        & --                  &  1.00 &                   0.74 &                      53 &    0.59(2) &    1.75(7) &     18.7(1.1) \\
                                                                       &         &            &           &                     &       &                        &                         &         &         &           \\
 $\mathtt{M0.23P0.73}^{\gamma = 7/5}_\mathrm{Cool: ff}\mathtt{A0.26}$  &     $512^3$ & 7/5        & free-free & 0.025               &  0.26 &                   0.75 &                     217 &    0.228(3) &    1.77(6) &     85.8(2.8) \\
 $\mathtt{M0.23P0.73}^{\gamma = 7/5}_\mathrm{Cool: lin}\mathtt{A0.26}$ &     $512^3$ & 7/5        & linear    & 0.018               &  0.26 &                   0.75 &                     217 &    0.226(3) &    1.77(6) &     86(3) \\
 $\mathtt{M0.35P0.72}^{\gamma = 7/5}_\mathrm{Cool: ff}\mathtt{A0.51}$  &     $512^3$ & 7/5        & free-free & 0.067               &  0.51 &                   0.75 &                      53 &    0.35(1) &    1.66(5) &     36.3(2.4) \\
 $\mathtt{M0.35P0.72}^{\gamma = 7/5}_\mathrm{Cool: lin}\mathtt{A0.51}$ &     $512^3$ & 7/5        & linear    & 0.049               &  0.51 &                   0.75 &                      53 &    0.35(1) &    1.67(6) &     37.1(2.2) \\
 $\mathtt{M0.50P0.74}^{\gamma = 7/5}_\mathrm{Cool: ff}\mathtt{A1.00H}$  &    $1024^3$ & 7/5        & free-free & 0.165               &  1.00 &                   0.71 &                      71 &    0.50(2) &    1.80(13) &     20.1(5) \\
 $\mathtt{M0.50P0.75}^{\gamma = 7/5}_\mathrm{Cool: lin}\mathtt{A1.00H}$ &    $1024^3$ & 7/5        & linear    & 0.125               &  1.00 &                   0.71 &                      71 &    0.50(2) &    1.74(8) &     20(1) \\
                                                                       &         &            &           &                     &       &                        &                         &         &         &           \\
 $\mathtt{M0.24P0.90}^{\gamma = 5/3}_\mathrm{Cool: ff}\mathtt{A0.39}$  &     $512^3$ & 5/3        & free-free & 0.051               &  0.39 &                   0.94 &                     272 &    0.24(1) &    1.98(8) &     81(4) \\
 $\mathtt{M0.23P0.72}^{\gamma = 5/3}_\mathrm{Cool: lin}\mathtt{A0.31}$ &     $512^3$ & 5/3        & linear    & 0.039               &  0.31 &                   0.75 &                     217 &    0.235(2) &    1.92(5) &     76.8(1.9) \\
 $\mathtt{M0.35P0.91}^{\gamma = 5/3}_\mathrm{Cool: ff}\mathtt{A0.73}$  &     $512^3$ & 5/3        & free-free & 0.132               &  0.73 &                   0.94 &                      67 &    0.35(1) &    1.80(10) &     37(3) \\
 $\mathtt{M0.36P0.72}^{\gamma = 5/3}_\mathrm{Cool: lin}\mathtt{A0.61}$ &     $512^3$ & 5/3        & linear    & 0.107               &  0.61 &                   0.75 &                      53 &    0.36(1) &    1.65(3) &     30.3(1.4) \\
 $\mathtt{M0.54P0.58}^{\gamma = 5/3}_\mathrm{Cool: ff}\mathtt{A1.00H}$  &    $1024^3$ & 5/3        & free-free & 0.250               &  1.00 &                   0.60 &                      72 &    0.54(2) &    1.83(6) &     16.8(7) \\
 $\mathtt{M0.54P0.57}^{\gamma = 5/3}_\mathrm{Cool: lin}\mathtt{A1.00H}$ &    $1024^3$ & 5/3        & linear    & 0.300               &  1.00 &                   0.60 &                      72 &    0.54(2) &    1.73(11) &     15.0(1.1) \\
                                                                       &         &            &           &                     &       &                        &                         &         &         &           \\
 $\mathtt{M0.37P1.14}^{\gamma = 5/3}_\mathrm{Cool: ff}\mathtt{A1.00}$  &     $512^3$ & 5/3        & free-free & 0.200               &  1.00 &                   1.40 &                     100 &    0.37(1) &    1.89(13) &     35.5(1.9) \\
 $\mathtt{M0.36P1.15}^{\gamma = 5/3}_\mathrm{Cool: ff}\mathtt{A1.00H}$  &    $1024^3$ & 5/3        & free-free & 0.200               &  1.00 &                   1.40 &                     100 &    0.36(1) &    1.72(8) &     31.2(1.5) \\
 $\mathtt{M0.41P0.91}^{\gamma = 5/3}_\mathrm{Cool: ff}\mathtt{A1.00}$  &     $512^3$ & 5/3        & free-free & 0.225               &  1.00 &                   1.20 &                      86 &    0.41(1) &    1.85(7) &     27.5(1.3) \\
 $\mathtt{M0.40P0.93}^{\gamma = 5/3}_\mathrm{Cool: ff}\mathtt{A1.00H}$  &    $1024^3$ & 5/3        & free-free & 0.225               &  1.00 &                   1.20 &                      86 &    0.40(1) &    1.75(9) &     24.9(1.5) \\
 $\mathtt{M0.46P0.73}^{\gamma = 5/3}_\mathrm{Cool: ff}\mathtt{A1.00}$  &     $512^3$ & 5/3        & free-free & 0.250               &  1.00 &                   1.00 &                      71 &    0.46(1) &    1.77(10) &     20.6(1.6) \\
 $\mathtt{M0.45P0.74}^{\gamma = 5/3}_\mathrm{Cool: ff}\mathtt{A1.00H}$  &    $1024^3$ & 5/3        & free-free & 0.250               &  1.00 &                   1.00 &                      71 &    0.45(1) &    1.7(10) &     19.5(1.4) \\
                                                                       &         &            &           &                     &       &                        &                         &         &         &           \\
 $\mathtt{M0.70P0.37}^{\gamma = 5/3}_\mathrm{Cool: ff}\mathtt{A1.00H}$    &    $1024^3$ & 5/3        & free-free & 0.330               &  1.00 &                   0.60 &                      43 &  0.70   &  1.66    &    9.81    \\
                                                                       &         &            &           &                     &       &                        &                         &         &         &           \\
 $\mathtt{M0.53P0.45}^{\gamma = 5/3}_\mathrm{Cool: ff}\mathtt{A0.80}$  &     $512^3$ & 5/3        & free-free & 0.211               &  0.80 &                   0.48 &                      34 &    0.53(2) &    1.68(9) &     15.0(7) \\
 $\mathtt{M0.48P0.58}^{\gamma = 5/3}_\mathrm{Cool: ff}\mathtt{A0.86}$  &     $512^3$ & 5/3        & free-free & 0.211               &  0.86 &                   0.60 &                      43 &    0.48(1) &    1.70(4) &     18.3(5) \\
 $\mathtt{M0.40P0.93}^{\gamma = 5/3}_\mathrm{Cool: ff}\mathtt{A1.00}$  &     $512^3$ & 5/3        & free-free & 0.211               &  1.00 &                   0.94 &                      67 &    0.40(1) &    1.78(3) &     26.4(7) \\
 $\mathtt{M0.52P0.59}^{\gamma = 5/3}_\mathrm{Cool: ff}\mathtt{A1.00}$  &     $512^3$ & 5/3        & free-free & 0.264               &  1.00 &                   0.60 &                      43 &    0.52(1) &    1.67(3) &     15.70(23) \\
 $\mathtt{M0.43P0.97}^{\gamma = 5/3}_\mathrm{Cool: ff}\mathtt{A1.16}$  &     $512^3$ & 5/3        & free-free & 0.264               &  1.16 &                   0.94 &                      67 &    0.43(1) &    1.81(7) &     23.6(1.1) \\
 $\mathtt{M0.50P1.01}^{\gamma = 5/3}_\mathrm{Cool: ff}\mathtt{A1.56}$  &     $512^3$ & 5/3        & free-free & 0.412               &  1.56 &                   0.94 &                      67 &    0.50(1) &    1.86(10) &     19.7(7) \\
 $\mathtt{M0.46P0.74}^{\gamma = 5/3}_\mathrm{Cool: lin}\mathtt{A1.00}$ &     $512^3$ & 5/3        & linear    & 0.220               &  1.00 &                   0.80 &                      57 &    0.46(1) &    1.74(6) &     20.5(7) \\
\bottomrule
\end{tabular}
\caption{Overview of the simulation parameters.
The simulations differ by the numerical resolution $N^3$,
the adiabatic index $\gamma$,
the cooling function (no, linear, or free-free) and cooling coefficient $C_\mathrm{cool}$,
the root mean square (RMS) power in the acceleration field $a$,
the initial thermal pressure $p_\mathrm{th,init}$, and
the initial ratio of thermal to magnetic pressure $\beta_\mathrm{p,init}$.
The saturated values of the 
and RMS sonic $\Ms$ and Alfv\'enic $\Ma$ Mach number and $\pb$
are calculated as the mean in the stationary regime between 
$5\mathrm{T} \leq t \leq 10\mathrm{T}$
(with temporal standard deviations in parentheses).
The values given for \PTI are the final values at $t=4\mathrm{T}$
when runaway cooling is triggered, see Sec.~\ref{sec:stability} for more details.
The simulation IDs \texttt{M\#\#\,P\#\#\,EOS\,A\#\#} are constructed using the
sonic Mach number and thermal pressure in the stationary regime, the EOS used, and
the forcing amplitude.
An ID ending with an \texttt{H} indicates a high resolution ($1024^3$) simulation
versus  $512^3$ without suffix.
}
\label{tab:overview}
\end{center}
\end{table*}

In total, we conduct 30 simulations.
All simulations evolve on a uniform, static, cubic grid with $512^3$ or $1024^3$ cells
and side length $L_{\mathrm{box}} = 1$ 
starting with uniform initial conditions (all in code units) $\rho = 1$, and
$\V{u} = \V{0}$.
The initial uniform pressure and background magnetic field (in the x-direction and
defined via the ratio of thermal to magnetic pressure, $\pb = \pth/\pB$)
vary between simulations as listed in Table~\ref{tab:overview}.
A regime of stationary turbulence is reached by a stochastic forcing
process that evolves in space and time so that no artificial compressive
modes are introduced to the simulation~\citep{Grete2018a}.
The forcing is purely solenoidal, i.e., $\nabla \cdot \V{a} = 0$, and the 
spectrum is parabolic with the peak at $k=2$ using
normalized wavenumbers \citep[see, e.g., ][for more details]{Schmidt2009}.
Thus, the characteristic length is $\mathrm{L} = 0.5L_\mathrm{box}$.
Given that all simulations reach a turbulent sonic Mach number of (or
close to)
$\Ms = u/\cs = 0.5$ in the stationary regime with speed of 
sound $\cs = \sqrt{\gamma \pth/\rho}$, 
we use $\cs$ as the characteristic velocity so that the 
dynamical time $\mathrm{T} = 1$.
Each simulations is evolved for 10~T, and 10 equally spaced snapshots per
dynamical time are stored for analysis.
The stationary regime is generally reached after approximately $3$T, and we 
exclude an additional $2$T as a few simulations required more time to reach equilibrium.
All statistical results presented below are only covering the
stationary regime, i.e., the statistics are calculated over 51 snapshots
between $5\mathrm{T} \leq t \leq 10\mathrm{T}$.

In general, the simulations can be separated along different parameter
dimensions in order to disentangle different competing effects with respect to 
parameters.
The main parameter dimensions target effects of different thermodynamics 
(i.e., the equation of state and cooling function) and with respect
to varying sonic Mach number $\Ms$.

With respect to thermodynamic effects, approximately isothermal runs
with $\gamma = 1.0001$ and no cooling are included as reference.
These are compared to simulations that employ different cooling functions
and ratios of specific heats, $\gamma$.
More specifically, $\gamma = 5/3$ (for a monoatomic gas) and 
$\gamma = 7/5$ (for a diatomic gas) are used, and cooling varies between
a linear cooling function and one that approximates free-free emission.
This results in sets of 5 simulations that differ in their thermodynamic properties.

In addition, these sets are compared at different (sub)sonic Mach numbers, which
is realized by either varying $u$ through varying forcing amplitudes and/or by 
varying the mean thermal pressure.
An overview of all simulations and their parameters is given in Table~\ref{tab:overview}.


\section{Results}
\label{sec:results}

\subsection{Overview}

All simulations reach an approximately stationary state in the 
subsonic ($0.2 \lesssim \Ms \lesssim 0.6$), 
super-Alfv\'enic ($\Ma  = u/\vA \approx 1.8$ 
with Alfv\'en velocity $\vA = B/\sqrt{\rho}$),
high $\pb$ ($10 \lesssim \pb \lesssim 100$) regime.
In other words, using these dimensionless numbers as proxies means 
that the kinetic energy, on average, is slightly lower 
than the thermal energy and slightly larger than the magetic energy,
and that the thermal energy (or pressure) is larger than the magnetic
energy (or pressure).

\begin{figure}[htbp]
\centering
\includegraphics{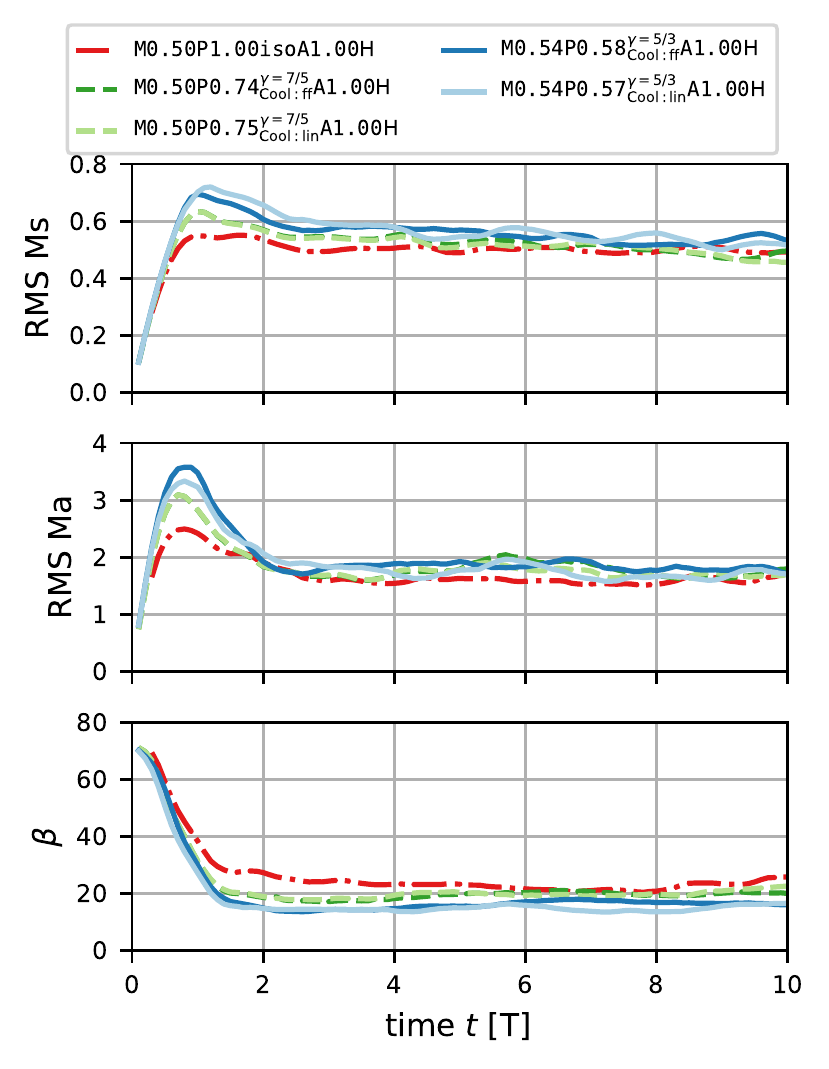}
\caption{Temporal evolution of the sonic Mach number $\Ms$,
the Alfv\`enic Mach number $\Ma$, and the ratio of thermal to
magnetic pressure $\pb$
for five simulations with $\Ms \approx 0.5$ in the stationary regime.
The simulations differ by the cooling function (none, linear, and free-free)
used and the adiabatic index $\gamma$.
}
\label{fig:evol-overview}
\end{figure}

For reference, the temporal evolution of $\Ms$, $\Ma$, and $\pb$ 
for five simulation with $\Ms \approx 0.5$ but with varying cooling function
and adiabatic index $\gamma$ is illustrated in Fig.~\ref{fig:evol-overview}.
The transient phase in which the initial conditions evolve towards
the stationary regime under constant driving lasts for about 3 dynamical times
(3~T).
In general, all data in the stationary regime presented in the following spans 
the temporal mean (and variations) between $5\mathrm{T} \leq t \leq 10\mathrm{T}$, which excludes 
an additional $2\mathrm{T}$ between $3\mathrm{T} \leq t \leq 5\mathrm{T}$ as few simulations took longer
to reach approximate equilibira.

Despite varying thermodynamics (isothermal EOS, adiabatic EOS with $\gamma = 5/3$ and 
$\gamma = 7/5$, and linear and free-free cooling) the five simulations in 
Fig.~\ref{fig:evol-overview} reach practically
identical $\Ms$, $\Ma$, and $\pb$ in the stationary regime (see 
Table~\ref{tab:overview}).

\begin{figure*}[htbp]
\centering
\includegraphics[width=\textwidth]{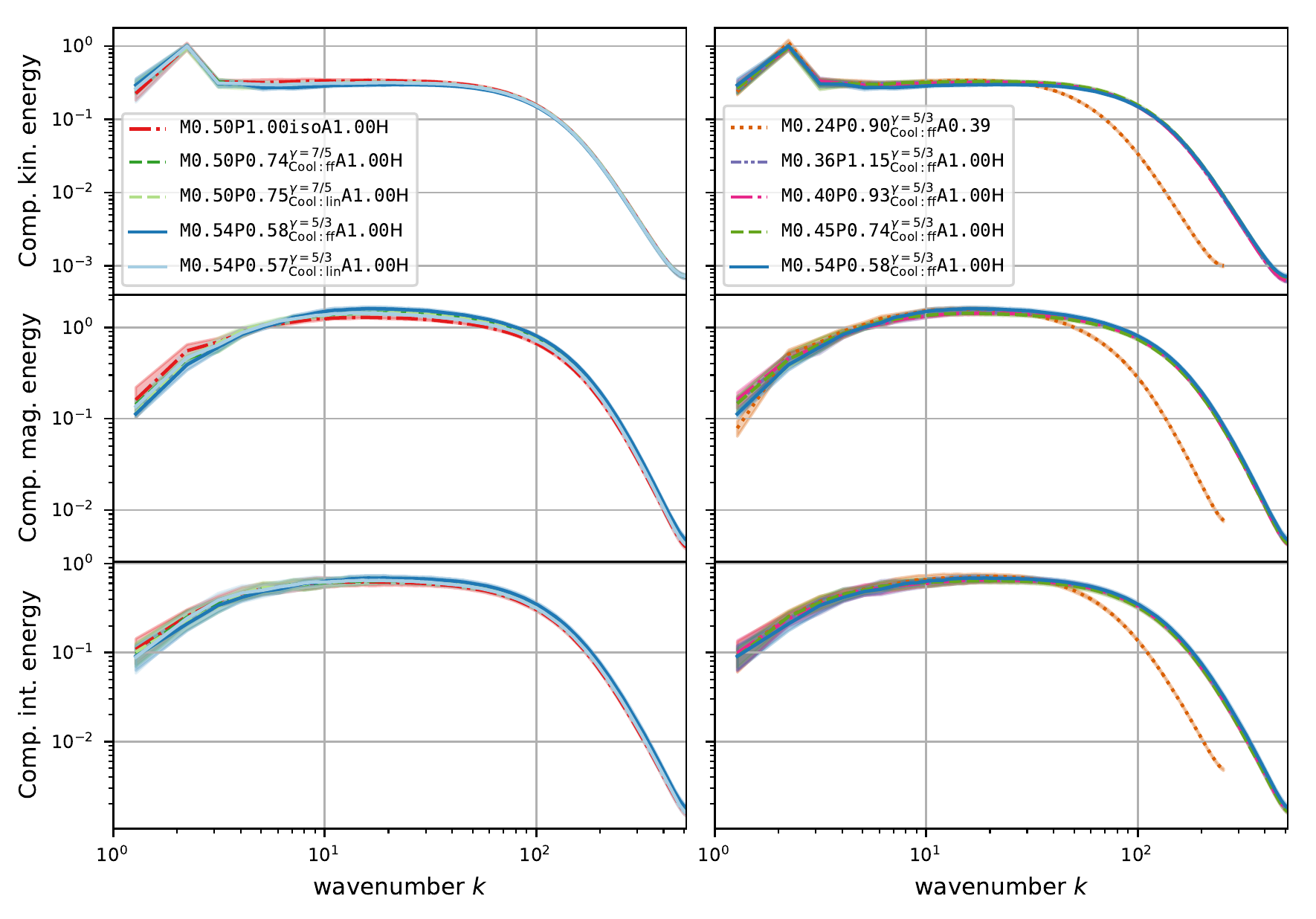}
\caption{
  Mean energy spectra of kinetic energy (top row),
  magnetic energy (middle row), and internal energy (bottom row).
  The mean is taken over the stationary regime $t > 5\mathrm{T}$ and the spectra
  are compensated by power laws of exponent 4/3, 5/3, and 4/3, respectively.
  The kinetic energy spectrum is calculated based on the Fourier 
  transform of $\sqrt{\rho}u$ and the internal energy spectrum
  based on $\sqrt{\rho}c_s$.
  All spectra are normalized to unit area under the curve.
  The left column shows the same simulations as in 
  Fig.~\ref{fig:evol-overview}, i.e., 
  $\Ms \approx 0.5$ with different cooling functions and the right column
  shows only
  simulations with free-free cooling and $\gamma = 5/3$ but
with varying $\Ms$.
Simulations with labels ending in \texttt{H} were run at $1024^3$ cell
resolution,
with all other calculations at $512^3$.
}
\label{fig:spec}
\end{figure*}
Similarly, the mean kinetic, magnetic, and internal energy spectra\footnote{
  The kinetic and internal energy spectra are calculated based on the Fourier transforms
  of $\sqrt{\rho}u$ and $\sqrt{\rho}c_s$.
  While this choice theoretically violates the inviscid criterion for decomposing scales
  for variable density flows \citep{Zhao2018}, we expect no practical differences for our
  simulations given the limited density variations in the subsonic regime.
}
are also identical as shown in Fig.~\ref{fig:spec} (left column) for the same 
five simulations.
The kinetic energy spectrum exhibits a power-law scaling within the
wavenumber range $4 \lesssim k \lesssim 40$.
No clear power-law scaling is observed in the magnetic and internal energy spectra.
The right column of Fig.~\ref{fig:spec} shows the energy spectra for five simulations
with free-free cooling and $\gamma = 5/3$ but with varying
$0.24 \lesssim \Ms \lesssim 0.54$.
Again, all spectra are identical between the simulations apart from the shorter
extent at high wavenumbers of the simulation run at $512^3$ compared to the others
run at $1024^3$.
General differences in the raw power (i.e., vertical offsets due to different numerical
values of, e.g., $u$ in the simulations) have been removed by normalizing the area 
under the spectra to unity.
This emphasizes the identical shape of the power spectra of all simulations.

\newpage

\subsection{Correlations}
\label{sec:corr}
\begin{figure*}[htbp]
\centering
\includegraphics[width=\textwidth]{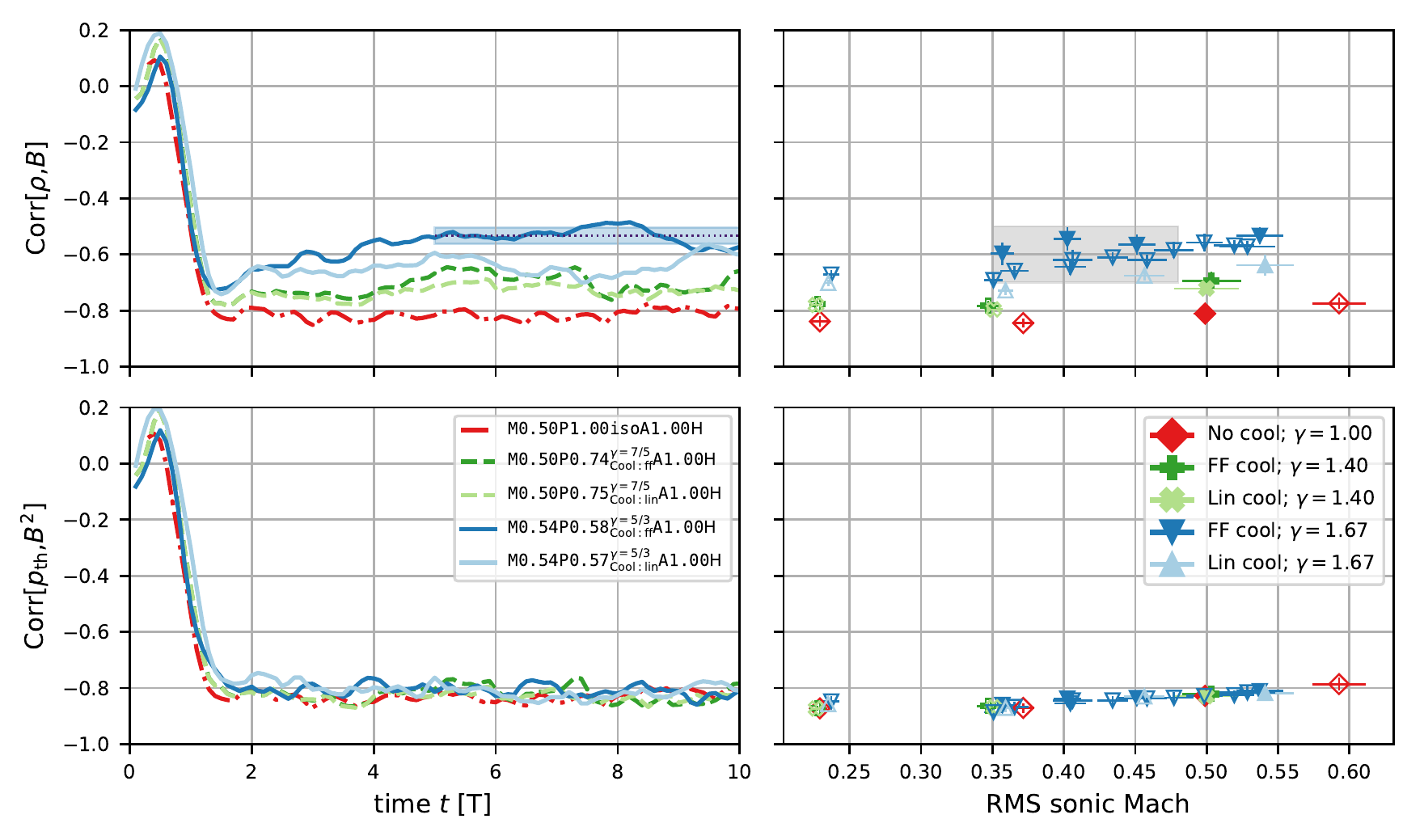}
\caption{
Correlation between
the density field $\rho$ and the magnetic field strength $B$ (top) and the
correlation between thermal and magnetic pressure (bottom).
The left column shows the temporal evolution of the correlations for the same
five simulations as in Fig.~\ref{fig:evol-overview}, i.e., simulations with 
$\Ms \approx 0.5$ but with varying EOS and cooling.
For the purpose of illustration, the dotted line in the top left panel indicates the mean 
value and the shaded area the standard deviation of that quantity 
over time ($5\mathrm{T}\leq t\leq10\mathrm{T}$) as it is used in the right column.
The right column shows the mean correlation coefficients 
versus sonic Mach number $\Ms$ in the stationary regime.
Each data point corresponds to one of 30 simulations total.
The horizontal and vertical lines for each symbol (usually within the bounds of the symbol)
correspond to standard deviation over time.
Filled symbols correspond to simulations run at a resolution of $1024^3$ and empty symbols
to a resolution of $512^3$.
The gray area highlights several simulations pairs running with identical parameters at
different resolution that are used for convergence analysis (see Appendix~\ref{sec:conv}).
  }
\label{fig:corr-param}
\end{figure*}
Similar to temporal evolution of the Mach numbers, the correlation coefficient between
thermal ($\pth$) and magnetic pressure ($\pB$) 
for the five simulations with $\Ms \approx 0.5$ and 
varying EOS and cooling settles to the same value of $\approx-0.8$ in the
stationary regime (see bottom left panel in Fig.~\ref{fig:corr-param}).
In contrast to this, different EOS and cooling functions result in different
correlation coefficients between the density field ($\rho$) and the magnetic field 
strength ($B$), as illustrated in the top left panel of Fig.~\ref{fig:corr-param}.
The isothermal reference case exhibits a strong anticorrelation of $-0.81(2)$ as 
previously observed in similar simulations \citep{Yoon2016,Grete2018a}.
The anticorrelation is weakened when departing from an isothermal equation of state.
For $\gamma=7/5$ the coefficient is $\approx -0.71$ independent of the 
cooling function, 
and for $\gamma=5/3$ it is $-0.64(3)$ in the case of linear cooling
and $-0.55(3)$ for free-free cooling.

These trends are also observed for different sonic Mach numbers, as shown in
the right column of Fig.~\ref{fig:corr-param}.
Here, the mean and standard deviation of the $\rho$--$B$ and $\pth-\pB$ correlation
coefficients in the stationary regime are illustrated versus sonic Mach number for
all 30 simulations.
In the regime presented, the $\pth-\pB$ correlation coefficient ($\approx-0.8$)
is practically independent of sonic Mach number, EOS, and cooling with a very weak 
trend towards weaker anticorrelation with increasing Mach number.
Overall, the thermal and magnetic pressure are highly anticorrelated.
This indicates a total pressure equilibrium 
(see also $\ptot$ distributions in the following Sec.~\ref{sec:pdfs}).

The individual $\rho$--$B$ correlation coefficients are predominately determined by
the EOS (here, via $\gamma$) and the cooling function used, see top right panel of
Fig.~\ref{fig:corr-param}.
A higher adiabatic index ($\gamma = 1.0001 \rightarrow 7/5 \rightarrow 5/3$) 
result in weaker anticorrelations. 
Moreover, free-free cooling results in slightly weaker anticorrelations compared to
linear cooling, but this effect is mostly visible in the $\gamma = 5/3$ simulations.
For example, for $\Ms \approx 0.35$ the $\rho$--$B$ correlation coefficients is
$-0.84$ in the isothermal case, $-0.80$ in the $\gamma = 7/5$ case with linear cooling
and $-0.78$ with free-free cooling, and $-0.73$ in case $\gamma = 5/3$ case with linear
cooling and $-0.69$ with free-free cooling.
Finally, there is an indication that higher numerical resolution also results
in slightly weaker $\rho$--$B$ anticorrelations (of about the same order as the cooling
functions).  Appendix~\ref{sec:conv} presentes a discussion of these
results with simulation resolution.

\subsection{Probability Density Functions}
\label{sec:pdfs}

\begin{figure*}[htbp]
\centering
\includegraphics[width=\textwidth]{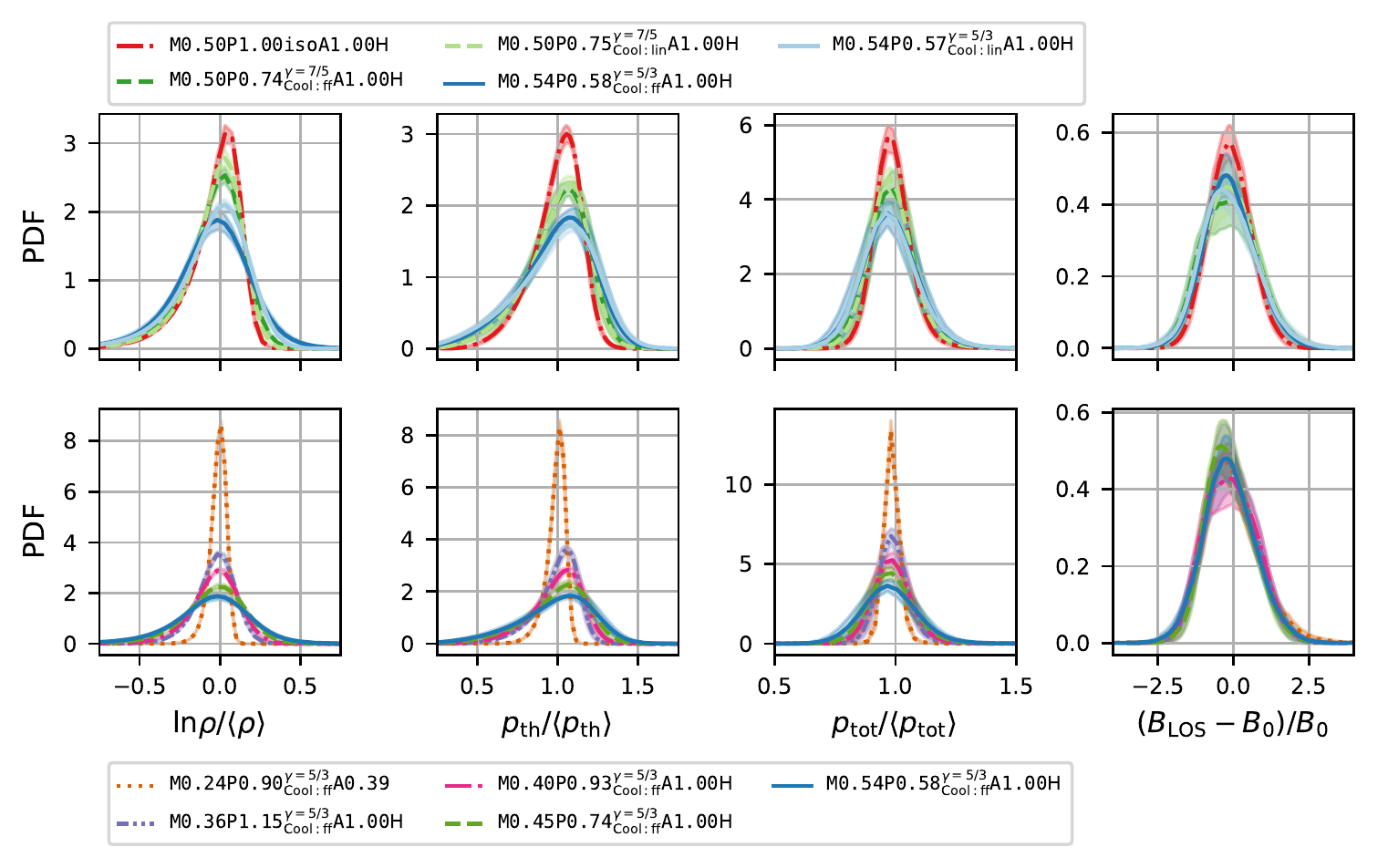}
\caption{
Mean probability density functions (PDF) of the density $\mathrm{ln}\rho$,
the normalized thermal pressure $\pth/\mean{\pth}$, the normalized total pressure
$\ptot/\mean{\ptot}$, and normalized deviation of the
derived line-of-sight magnetic field strength to the actual one
$(B_\mathrm{LOS} - B_0)/B_0$.
Normalization is applied with respect to the mean value.
The mean is taken over the stationary regime ($t > 5\mathrm{T}$) and the shaded regions
indicate the standard deviation of the PDFs over time.
The rows show the same simulation as in Fig.~\ref{fig:evol-overview}, i.e., 
$\Ms \approx 0.5$ with different cooling functions.
The bottom row depicts only simulations with free-free cooling and $\gamma = 5/3$ but
with varying resolution, $\Ms$, and power in the forcing field.
}
\label{fig:hist}
\end{figure*}
Similar to the $\rho$--$B$ correlation coefficients different $\gamma$ and different
cooling functions lead to systematically changing statistics in other quantities.
Figure~\ref{fig:hist} shows the mean probability density functions (PDFs) of the
density $\mathrm{ln}\rho$,
the normalized thermal pressure $\pth/\mean{\pth}$, the normalized total pressure
$\ptot/\mean{\ptot}$, and normalized deviation of the
derived line-of-sight (LOS) magnetic field strength to the actual one
$(B_\mathrm{LOS} - B_0)/B_0$.
The latter is derived from rotation measures\footnote{
  In principle, this relation holds for the number density of thermal electrons,
  but given the single fluid MHD approximation the
  density $\rho$ is used instead.}
via
\begin{eqnarray}
\label{eq:LOSB}
B_\mathrm{LOS} = \frac{\int_L \rho\bra{l} B_\parallel \bra{l} \mathrm{d} l }
{\int_L \rho\bra{l} \mathrm{d} l}
\end{eqnarray}
with $B_\parallel$ being the line-of-sight component of the magnetic field.

The top row in Fig.~\ref{fig:hist} shows five simulations with $\Ms \approx 0.5$ and
with different EOS and cooling functions.
Differences in the shape of the PDFs of the density, thermal pressure and total pressure
between the isothermal case,  $\gamma = 7/5$, and $\gamma = 5/3$ are immediately apparent.
For example, with higher $\gamma$ the PDF of $\mathrm{ln}\rho$ becomes less skewed and
all three PDFs become broader.
Differences between linear and free-free cooling are more subtle as discussed below.
No clear signal between different EOS is observed in the derived LOS magnetic 
field strengths.

The bottom row in Fig.~\ref{fig:hist} shows five simulation with $\gamma = 5/3$ and 
free-free cooling but with varying sonic Mach number ($0.25 \lesssim \Ms \lesssim 0.55$).
Again, differences in the shapes of the PDFs of the density, thermal pressure, and 
total pressure are apparent.
With increasing sonic Mach number all PDFs become broader.
In addition, the PDF of the thermal pressure becomes less skewed with with increasing $\Ms$
while the PDF of the total pressure remains mostly symmetric.
In fact, these differences with $\Ms$ are much more pronounced (cf., the scaling of the 
y-axis), suggesting that the sonic Mach number is the dominant effect compared to
changes in the EOS and cooling.
Again, no significant differences in the PDFs of the LOS magnetic field strength
are observed.

\begin{figure*}[htbp]
\centering
\includegraphics[width=\textwidth]{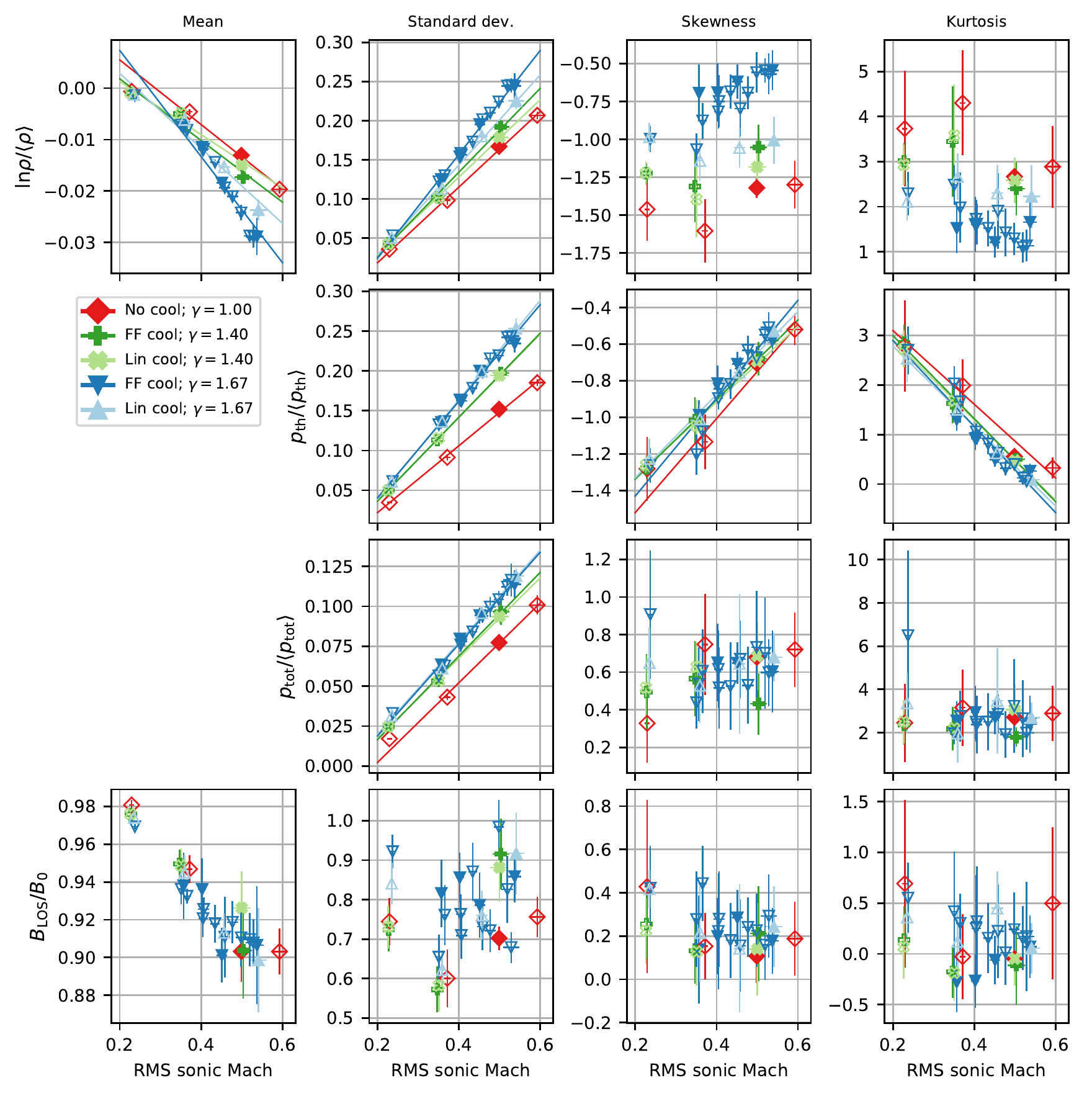}
\caption{
Statistical moments (from left to right mean, standard deviation, skewness, and
kurtosis) of the density, thermal pressure, total pressure, and derived magnetic
field strength (top to bottom) versus sonic Mach number $\Ms$ in the stationary regime.
Each data point corresponds to one of 30 simulations total.
The horizontal and vertical lines for each symbol (usually within the bounds of the symbol)
correspond to standard deviation over time.
Filled symbols correspond to simulations run at a resolution of $1024^3$ and empty symbols
to a resolution of $512^3$.
The lines in the panels of $\mathrm{ln}\rho$ (mean and std.~dev.), thermal pressure 
(all panels), and total pressure (std. dev.) indicate linear fits with $\Ms$.
Slope and correlation coefficients are given in Table~\ref{tab:fit}.
  }
\label{fig:stats-vs-Ms}
\end{figure*}

In order to further quantify the results, we calculate the statistical moments (mean,
standard deviation, skewness, and kurtosis) of these four quantities for all snapshots
of all 30 simulations.
The skewness is calculated as
\begin{equation}
  \Skew{x} = \frac{\left < \bra{x - \left < x \right > }^3 \right > }{\sigma^3 \bra{x}}
\end{equation}
with standard deviation $\sigma$ and the (Fisher) kurtosis is calculated as
\begin{equation}
  \Kurt{x} = \frac{\left < \bra{x - \left < x \right > }^4 \right > }{\sigma^4 \bra{x}} -3 \;. 
\end{equation}
The mean (over time) statistical moments including their standard deviations (over time)
versus sonic Mach number $\Ms$ are illustrated in Fig.~\ref{fig:stats-vs-Ms}.
Moreover, in cases where absolute correlation between a quantity $x$ and $\Ms$ is
larger than 0.9 we perform a linear fit with
\begin{equation}
x = m \Ms + b \,.
\end{equation}
The regressions are done over all outputs of all simulations employing a particular
combination of EOS and cooling, e.g., for $\gamma=7/5$ with linear cooling 
$3\times51=153$ data points are taken into account or $4\times51=204$ in the
isothermal case.
The slope $m$ of the fit and the correlation coefficient are given in Table~\ref{tab:fit}.
Note that given the limited range of $\Ms$ these fits need to be interpreted with care
and we primarily use them here in order to quantify differences (or the absence thereof)
between different equations of state and cooling functions.

Both the mean and the standard deviation $\sigma_{\mathrm{ln}\rho}$ exhibit a high 
(anti)correlation with $\Ms$ of $\leq -0.95$ and $\geq0.98$, respectively.
The trend of broader $\mathrm{ln}\rho$ distributions with increasing $\Ms$ observed
in Fig.~\ref{fig:hist} holds across all combinations of EOS and cooling.
Based on the slopes this trend is more pronounced both with larger $\gamma$ and with 
cooling that is more sensitive to $\rho$.
In the isothermal reference case the slope is shallower (-0.48) compared to
$\gamma=7/5$ with linear (-0.50) and free-free (-0.54) cooling, and to $\gamma=5/3$
with linear (-0.58) and free-free (0.66) cooling.
For the skewness and kurtosis no dependency on $\Ms$ is observed.
However, the distributions generally separate for different $\gamma$ and become less
skewed and less broad with increasing $\gamma$.

All statistical moments of the normalized thermal pressure 
(second row in Fig.~\ref{fig:stats-vs-Ms}) vary with $\Ms$.
Similar to the density distributions, the standard deviation of the thermal pressure
tightly depends on $\Ms$ (correlation coefficient $\geq0.98)$ and 
exhibits an additional (weaker) dependency on $\gamma$. 
With increasing $\gamma$ the slopes are getting steeper, from $\approx0.42$ in the
isothermal case, to $\approx0.52$ for $\gamma=7/5$, to $\approx0.61$ for $\gamma=5/3$
(with no pronounced difference between cooling functions).
For the skewness and kurtosis the correlations with $\Ms$ are generally weaker but still
pronounced ($\geq0.86$) and a clear trend differentiating EOSs and cooling functions is
not observed.

In contrast to this, the skewness and kurtosis of the normalized total pressure 
distributions (third row in Fig.~\ref{fig:stats-vs-Ms}) are independent of $\Ms$ and
also independent of $\gamma$ or cooling function.
However, the standard deviation is again tightly correlated ($\geq0.97$) with $\Ms$ 
and, similarly to the thermal pressure and density, shows an additional (weaker)
dependency on $\gamma$ with steeper slopes for larger $\gamma$.

Finally, the statistical moments of the 
normalized derived line-of-sight magnetic field strength distributions
(bottom row in Fig.~\ref{fig:stats-vs-Ms}) generally exhibit no clear trend with  
$\Ms$,  EOS, or cooling.
A weak trend is seen only in the mean value for a stronger 
underestimation of the field strength with increasing $\Ms$, but the scatter is too large
to large to make a definite statement.

\begin{table*}
\begin{center}
\begin{tabular}{lllccccc}
\toprule
Quan. & Stat. & & isoth. & $\gamma=7/5$ lin. & $\gamma=7/5$ ff. & $\gamma=5/3$ lin. & $\gamma=5/3$ ff. 
\\
\midrule
\multirow{2}{*}{$\mathrm{ln}\rho/\mean{\rho}$} & \multirow{2}{*}{mean}  
  & $m$  & -0.063(1) & -0.052(1) & -0.060(1) & -0.073(1) & -0.103(1) \\
& & Corr & -0.98     & -0.98     & -0.97     & -0.97     & -0.95 \\
\\
\multirow{2}{*}{$\mathrm{ln}\rho/\mean{\rho}$} & \multirow{2}{*}{std.}  
  & $m$  & 0.483(7)  & 0.495(5)  & 0.536(6)  & 0.575(6)  & 0.664(5)  \\
& & Corr & 0.99      & 0.99      & 0.99      & 0.99      & 0.98  \\
\\
\multirow{2}{*}{$\pth/\mean{\pth}$} & \multirow{2}{*}{std.}  
  & $m$  & 0.420(6)  & 0.523(4)  & 0.528(5)  & 0.624(5)  & 0.607(4)  \\
& & Corr & 0.99      & 0.99      & 0.99      & 0.99      & 0.98  \\
\\
\multirow{2}{*}{$\pth$} & \multirow{2}{*}{skew} 
  & $m$  & 2.57(10)  & 2.07(7)   & 2.18(7)   & 2.27(5)   & 2.68(5)   \\
& & Corr & 0.90      & 0.92      & 0.94      & 0.95      & 0.89  \\
\\
\multirow{2}{*}{$\pth$} & \multirow{2}{*}{kurt.} 
  & $m$  & -7.4(4)   & -8.1(3)   & -8.4(3)   & -8.1(2)   & -8.7(2)   \\
& & Corr & 0.86      & 0.93      & 0.94      & 0.95      & 0.91  \\
\\
\multirow{2}{*}{$\ptot/\mean{\ptot}$} & \multirow{2}{*}{std.} 
& $m$  & 0.250(4)  & 0.252(2)  & 0.262(3)  & 0.289(4)  & 0.288(3)  \\
& & Corr & 0.98      & 0.99      & 0.99      & 0.99      & 0.97  \\
\bottomrule
\end{tabular}
\caption{Overview of the numerical values of the linear fitting results illustrated in
Fig.~\ref{fig:stats-vs-Ms}. $m$ is the slope of the linear fit 
(including standard deviation) and Corr the correlation coefficient. 
}
\label{tab:fit}
\end{center}
\end{table*}

\subsection{Pressure--density dynamics and thermal stability}
\label{sec:stability}
\begin{figure*}[htbp]
\centering
\includegraphics[width=\textwidth]{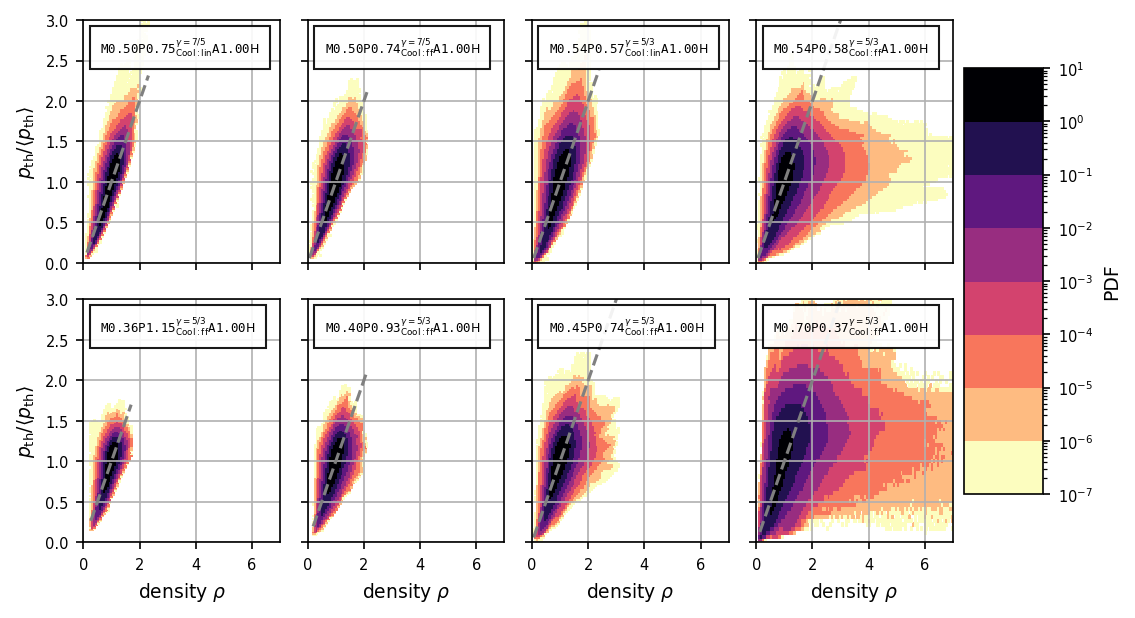}
\caption{Mean 2D PDFs of the normalized thermal pressure $\pth/\mean{p_{\mathrm{th}}}$ 
versus density.
The mean is covering 51 snapshots in in the stationary regime $5\mathrm{T} \leq t \leq 10\mathrm{T}$.
For reference, the gray dashed lines indicates $\pth/\mean{\pth} = \rho$.
The top row illustrates the same simulations as in Fig.~\ref{fig:evol-overview},
i.e., $\Ms \approx 0.5$ with 
different cooling functions, and the bottom row shows simulations with
free-free cooling and $\gamma = 5/3$ but varying $\Ms$ (through lowering the mean pressure
from left to right).
Instead of the mean PDF the bottom right figure shows the final PDF at $t=4\mathrm{T}$ 
when runaway cooling is triggered, see Sec.~\ref{sec:stability} for more details.
}
\label{fig:hist-prho}
\end{figure*}

In order to first understand the individual distributions presented in the
previous section, the mean 2D PDFs of thermal pressure versus density are
illustrated in Fig.~\ref{fig:hist-prho}.

The top row depicts the PDFs for the simulations at $\Ms\approx0.5$ with
varying $\gamma$ and cooling.
In general, the distributions are extended around the isothermal reference line
($p\propto\rho$) as expected given the chosen balance between turbulent dissipation
and cooling.
With higher $\gamma$ the distributions are getting broader in both dimensions.
Moreover, free-free cooling leads to an additional broadening in the density
dimension in both cases for $\gamma=7/5$ and $\gamma=5/3$.
The most extreme case ($\gamma = 5/3$ with free-free cooling in the top right panel)
exhibits broad density tails as the simulation is thermally marginally stable.

To further illustrate the transition to a thermally unstable regime the
bottom row in Fig.~\ref{fig:hist-prho} shows the mean 2D PDFs only for simulations
with $\gamma = 5/3$ and free-free cooling but with increasing $\Ms$ (going from
$0.36$ to $\approx0.7$).
The increasing $\Ms$ (for the same forcing amplitude) is achieved by lowering the 
mean thermal pressure in the simulations.
With increasing sonic Mach number the distributions are getting broader in
both dimensions.
This can be attributed to the increasing width of the density PDF with $\Ms$ 
(see Sec.~\ref{sec:pdfs}), which is enabled by decreasing pressure support
against compression.
The bottom right panel shows simulation \PTI for which the pressure support
is insufficient to prevent runaway cooling resulting in extended high density tails.

\section{Discussion}
\label{sec:disc}
\subsection{Correlations and relevance to observations}
The correlation between the density $\rho$ and magnetic field strength 
$B$ is astrophysically relevant to the line-of-sight (LOS) magnetic field 
strength measurement via Faraday rotation.
Only for uncorrelated fields and in the isothermal case
the derived strength is exact \citep{Beck2013}.
Here, we observe that the $\rho$--$B$ correlation depends on the adiabatic
index $\gamma$, i.e., there is a clear departure from the isothermal case.
The strong anticorrelation observed in isothermal simulation weakens with larger
$\gamma$.
For isothermal simulations it was additionally observed that the correlation depends on the
sonic Mach number $\Ms$ (especially when going to the supersonic regime) and
the correlation time of the forcing \citep{Yoon2016,Grete2018a,Beckwith2019}.
However, the mean deviation of the derived LOS magnetic field from the exact one
in our simulations is at most 10\%, with a trend of the deviation
becoming more significant going
from $\Ms \approx 0.2$ to $\approx 0.6$ independent of different thermodynamics.
Thus, the resulting deviation for ICM-like plasmas is likely below the 
observational uncertainties.

Overall, $\rho$--$B$ are anticorrelated across all sonic Mach numbers presented.
For isothermal MHD \citet{Passot2003} showed that this anticorrelation is indicative of 
dynamics governed by slow magnetosonic modes.
The dependency on $\gamma$ in the $\rho$--$B$ correlation observed in this paper suggests
that there exists a richer mix of modes when departing from an isothermal equation of 
state.

In contrast to the $\rho$--$B$ correlation, the correlation between thermal and magnetic
pressure is independent of $\gamma$ and cooling, i.e., the correlation coefficient
of the isothermal simulation is indistinguishable from the adiabatic simulations with 
cooling.
This suggests that all simulations are governed by a total pressure equilibrium,
$\pth + \pB = \ptot \approx \mathrm{const.}$.

\subsection{$\sigma_\rho$--$\Ms$ relation and comparison to previous work}
The presented work covers isothermal and non-isothermal magnetized stationary turbulence
with varying $\gamma$ and cooling functions over a range of $\Ms$ in the subsonic regime.
The majority of previous related work comes from the star formation community and
targets isothermal, (magneto)hydrodynamic, supersonic turbulence.

Initial work on the relation between density variations and the sonic Mach number
goes back to \citet{Padoan1997,Passot1998} who derived and tested numerically the linear
relation
\begin{equation}
\sigma_{\rho/\mean{\rho}} = b\Ms\,.
\end{equation}
In the case of isothermal hydrodynamic turbulence, \citet{Federrath2008} later showed
that the proportionality constant $b$ varies depending on the modes employed in the
forcing (between $\approx0.3$ for purely solenoidal forcing and $\approx1$ for purely
compressive forcing).

Qualitatively, we also find a linear relation between density variations\footnote{
  Note that the slopes and correlations of the fit reported in Table~\ref{tab:fit} are for
  $\ln (\rho/\mean{\rho})$ instead of $\rho/\mean{\rho}$, but we find similar behavior
  (i.e., a linear relation) for the latter.}
and $\Ms$.
Moreover, we find that the slope of the linear relation depends on both the adiabatic index
$\gamma$ (steeper with larger $\gamma$) 
and the cooling employed for identical, purely solenoidal forcing.
This suggests that departure from an isothermal regime adds additional complexity to the
relation, which is also found by \citet{Nolan2015} in the hydrodynamic case.
The latter presents both numerical results and a theoretical model that predicts steeper
slopes for larger $\gamma$ in the subsonic regime.

In the MHD case adjustments to the relation have been reported 
\citep[by, e.g.,][]{Padoan2011,Molina2012} that take the ratio of thermal to magnetic
pressure, $\pb$, into account.
However, the adjustment is of the order of $1/\sqrt{1+\pb^{-1}}$.
Given that  $10\lesssim\pb \lesssim100$ in all of our simulations, this correction would
contribute at most a few percent to our results and is thus negligible.

\citet{Kowal2007} studied  density fluctuations in isothermal MHD turbulence, 
including higher order statistical moments such as skewness and kurtosis.
However, the random (uncorrelated) forcing employed in \citet{Kowal2007} leads to
the excitation of compressive modes (despite solenoidal forcing) in the subsonic regime 
that systematically affect several statistics including the correlation between density 
and magnetic field strength or the density PDF \citep{Yoon2016,Grete2018a}.
This systematic effect renders a direct comparison difficult as the results are 
dependent on multiple parameters.
For example, a shorter autocorrelation in the forcing requires a larger forcing amplitude
to reach the same sonic Mach number resulting in more power in compressive modes.
In turn, this changes the density PDF for identical sonic Mach numbers and, thus, is
complementary to the changes described for varying Mach number and EOS in this manuscript.

Finally, it should be noted that our results are not in agreement with recently published
results by \citet{Mohapatra2019} who conduct adiabatic hydrodynamic simulations 
with and without heating/cooling.
They find $\sigma_{\pth/\mean{\pth}} \propto \sigma_{\rho/\mean{\rho}} \propto \Ms^2$
in the subsonic regime without cooling, and 
$\sigma_{\pth/\mean{\pth}} \propto \Ms^2$  and $\sigma_{\rho/\mean{\rho}} > \Ms$ with 
cooling whereas we find linear relationships for both density and pressure fluctuations.
Given the differences in the setup, e.g., MHD versus HD, idealized cooling versus 
realistic cooling curve, thermally unstable versus stable,
and heating only via turbulent dissipation versus turbulent dissipation and 
explicit heating, the observed differences in the results may stem from a variety of
sources or a combination thereof. As a result, we refrain from a more detailed, purely
speculative comparison between the results presented here and those of \citet{Mohapatra2019}.

\subsection{Limitations}
Given the idealized nature of this work several items need to be kept in mind when
interpreting or extrapolating from the results.

This pertains, for example, to the idealized cooling functions that in their current
form only approximate subregimes of a realistic cooling function.
Similarly, given the monotonic shape of the cooling functions and the targeted
balance between turbulent dissipation and cooling (to achieve stationary turbulence)
prevents the development of multi-phase flows.
Thus, the results presented provide a qualitative view on the effects of different
cooling functions and equation of states. 
For detailed predictions in specific environments such as different phases
in the ISM more realistic cooling functions should be employed.

In addition, the sampled parameter space is mostly targeted at ICM-like regimes, i.e.,
subsonic, super-Alfv\'enic, and high $\pb$ turbulence, though neglecting effects from
low collisionality in the ICM that can also alter statistical moments of, for example,
the density distribution \citep{Schekochihin2006,Kowal2011}.
While several clear trends in the $\rho$--$B$ correlations with varying
thermodynamics and in the distribution functions of $\rho$, $\pth$, and $\pB$ with
varying $\Ms$ have been observed, the resulting relations should be handled with care
-- especially in extrapolating to the supersonic regime.

\section{Conclusions}
\label{sec:conclusions}
In this paper, we systematically studied how the departure from an isothermal
equation of state affects stationary magnetohydrodynamic turbulence.
In total, we conducted 30 numerical simulations with varying adiabatic index
$\gamma$ with $\gamma\rightarrow1$ for an approximately isothermal gas as reference case,
$\gamma=7/5$ for a diatomic gas, and $\gamma=5/3$ for a monoatomic gas.
Moreover, we employed two idealized cooling function (linear cooling with $\dot E \propto \rho e$ and approximate free-free emission with $\dot E \propto \rho^2 \sqrt{e}$)
in order to maintain stationary turbulence with a constant Mach number.
All simulations are subsonic ($\Ms \approx 0.2$ to $0.6$), 
super-Alfv\'enic ($\Ma \approx 1.8$), and high $\pb$ (ratio of
thermal to magnetic pressure with $10 \lesssim \pb \lesssim 100$) -- a
regime found, 
for example, in the intracluster medium.

In this regime, we find that the kinetic, magnetic, and internal energy spectra are 
practically unaffected by the thermodynamics and the sonic Mach number (apart from the normalization).
Moreover, the thermal and magnetic pressures are strongly anticorrelated (correlation 
coefficient $\lesssim-0.8$) independent of $\gamma$ and cooling, and only exhibit
a weak trend towards weaker anticorrelation with increasing $\Ms$.
In contrast to this, the correlation between density $\rho$ and magnetic field strength, 
$B$ (which, again, are anticorrelated) shows a dependency on $\gamma$.
The correlation coefficient of $\approx-0.8$ in the isothermal reference case gets weaker
with larger $\gamma$ up to $\approx-0.55$ for $\gamma=5/3$ with free-free cooling.
Departing from an isothermal equation of state allows independent
thermal pressure and density variations.
Thus, for a fixed, strong $\pth$--$\pB$ anticorrelation (associated with a total pressure
equilibrium) and adiabatic equation of state naturally reduces the 
$\rho$--$B$ correlation by construction.

Similarly, we find dependencies on $\gamma$ in multiple distribution functions.
However, these dependencies are typically subdominant with respect to the overall
trend with $\Ms$.
For example, we find linear relations for an increase of density fluctuations, 
thermal and total pressure fluctuations, and the skewness of the thermal pressure
distribution with increasing $\Ms$.
A larger $\gamma$ and a cooling function with stronger density dependency generally 
result in slightly steeper slopes of these linear relations.

Overall, this results in degeneracy in inferring, for example,
Mach numbers from observed distributions without knowing the governing thermodynamics
of the observed system.
However, we suggest that higher order statistics (e.g., the skewness
of the density distribution) are less dependent on $\Ms$ and predominately determined
by $\gamma$.
Thus, there is hope that this degeneracy can be resolved.
To do so would require simulations that span a substantially broader
and more complex parameter space,  which
we leave to future work.

\acknowledgments
The authors thank Wolfram Schmidt and Christoph Federrath for useful discussions.
PG and BWO acknowledge funding by NASA Astrophysics Theory Program
grant \#NNX15AP39G.
Sandia National Laboratories is a multimission laboratory managed and operated by 
National Technology and Engineering Solutions of Sandia LLC, a wholly owned 
subsidiary of Honeywell International Inc., for the U.S. Department of Energy's 
National Nuclear Security Administration under contract DE-NA0003525.
This paper describes objective technical results and analysis.
Any subjective views or 
opinions that might be expressed in the paper do not necessarily represent the 
views of the U.S. Department of Energy or the United States Government.
BWO acknowledges additional funding by NSF AAG grant \#1514700.
The simulations were run on the NASA Pleiades supercomputer through allocation SMD-16-7720 
and on the Comet supercomputer as part of the Extreme Science and Engineering Discovery
Environment \citep[XSEDE][]{XSEDE}, which is supported by National Science Foundation 
grant number ACI-1548562, through allocation \#TG-AST090040.
\textsc{Athena} is developed by a large number of independent
researchers from numerous institutions around the world. Their
commitment to open science has helped make this work possible.
SAND Number: SAND2019-14747 J

\software{
  Modified version of \texttt{Athena} \citep[v4.2][]{Stone2008} available
  at \url{https://github.com/pgrete/Athena-Cversion} (\texttt{3a7c300}).
  \texttt{Matplotlib} \citep{matplotlib}.
  \texttt{NumPy} \citep{numpy}. 
  \texttt{mpi4py} \citep{mpi4py}.
  \texttt{mpi4py-fft} \citep{mpi4py-fft}.
}

\bibliographystyle{yahapj}

\appendix
\section{Convergence of simulations}
\label{sec:conv}
\begin{figure*}[htbp]
\centering
\includegraphics[width=\textwidth]{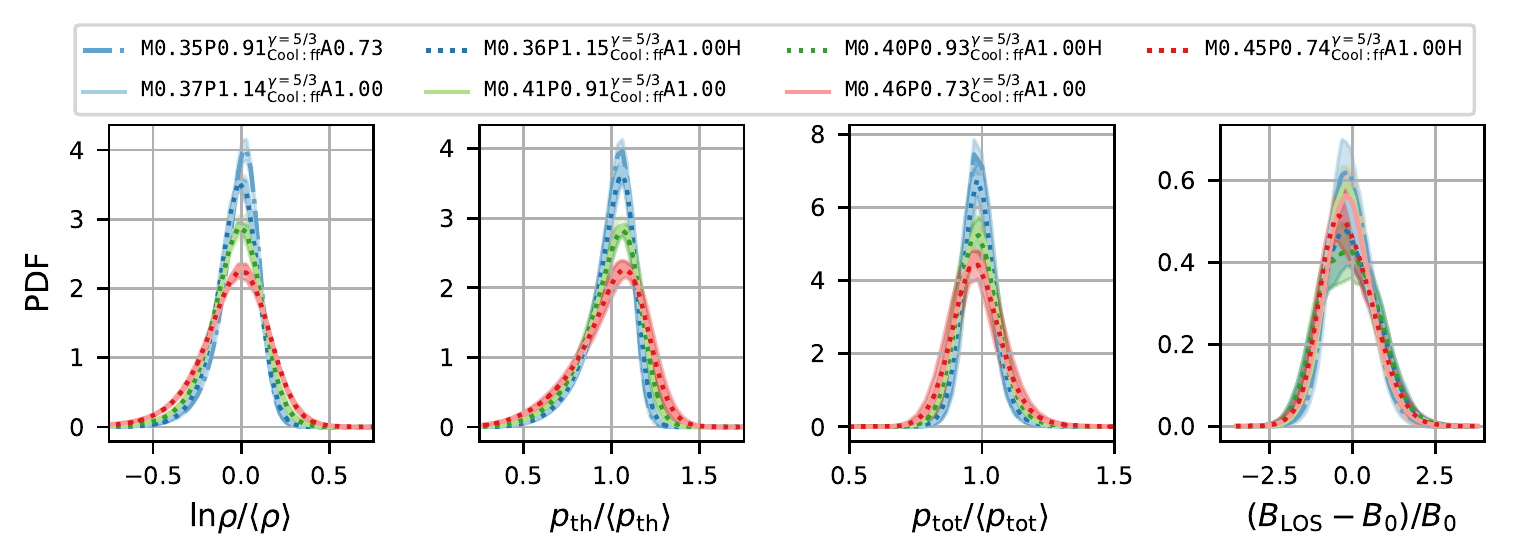}
\caption{
Mean probability density functions (PDF) of the density $\mathrm{ln}\rho$,
the normalized thermal pressure $\pth/\mean{\pth}$, the normalized total pressure
$\ptot/\mean{ptot}$, and normalized deviation of the
derived line-of-sight magnetic field strength to the actual one
$(B_\mathrm{LOS} - B_0)/B_0$.
Normalization is applied with respect to the mean value.
The mean is taken over the stationary regime ($t > 5\mathrm{T}$) and the shaded regions
indicate the standard deviation of the PDFs over time.
All simulations use free-free cooling and $\gamma=5/3$ and are but vary (in pairs with 
same $\Ms$) in numerical resolution plus one additional simulation (at $\Ms \approx 0.35$)
with different forcing amplitude.
Each panel shows all 7 simulation and the lines for simulations with same $\Ms$ 
(i.e., same color) but different resolution ($512^3$ solid lines and $1024^3$ dashed lines)
are on top of each other illustrating convergence of the PDFs with resolution.
}
\label{fig:hist_appendix}
\end{figure*}
All simulations presented in this paper were conducted at a grid
resolution of either $512^3$ or $1024^3$ 
grid cells, with the latter indicted by a $\mathtt{H}$ suffix in the simulation ID.
While small differences between simulations with identical parameters but different
resolutions were observed in the correlations (see Sec.~\ref{sec:corr}), the
dominating effect determining the statistics discussed in the paper is related
to varying sonic Mach number.
Moreover, the PDFs are converged with resolution as illustrated in 
Fig.~\ref{fig:hist_appendix} where three sets of simulations with identical $\Ms$ 
($\approx0.36,\,0.4,\,\mathrm{and}\,0.45$) for both resolutions are shown.
In general, the PDFs for the same $\Ms$ (i.e., same color) are on top of each other, i.e.,
the solid lines for simulations at $1024^3$ are below the dotted lines of simulations
at $512^3$, illustrating convergence.

\end{document}